\documentclass[11pt]{article}

\usepackage[utf8]{inputenc}
\usepackage[T1]{fontenc}
\usepackage{amsmath,amssymb,amsfonts}
\usepackage{graphicx}
\usepackage{xcolor}
\usepackage{hyperref}
\usepackage{natbib}
\usepackage{booktabs}
\usepackage{subfig}
\usepackage[margin=1in]{geometry}
\usepackage{listings}
\usepackage[colorinlistoftodos]{todonotes}
\usepackage{authblk}

\definecolor{lightblue}{rgb}{0.8,0.9,1}
\presetkeys{todonotes}{inline,backgroundcolor=lightblue}{}

\hypersetup{
    colorlinks=true,
    linkcolor=blue,
    filecolor=magenta,      
    urlcolor=cyan,
    citecolor=blue,
    pdftitle={Towards the Autonomous Optimization of Urban Logistics: Training Generative AI with Scientific Tools via Agentic Digital Twins and Model Context Protocol},
    pdfauthor={Haowen Xu, Yulin Sun, Jose Tupayachi, Olufemi Omitaomu, Sisi Zlatanova, Xueping Li}
}

\title{Towards the Autonomous Optimization of Urban Logistics: Training Generative AI with Scientific Tools via Agentic Digital Twins and Model Context Protocol}

\author[1,2]{\small Haowen Xu}
\author[3]{\small Yulin Sun}
\author[4]{\small Jose Tupayachi}
\author[1]{\small Olufemi Omitaomu}
\author[2]{\small Sisi Zlatanova}
\author[4]{\small Xueping Li\thanks{Corresponding author: Xueping.Li@utk.edu}}

\affil[1]{\small Oak Ridge National Laboratory, 1 Bethel Valley Rd, Oak Ridge, TN 37830, USA}
\affil[2]{\small GRID, School of Built Environment, UNSW Sydney, NSW 2052, Australia}
\affil[3]{\small Southwestern University of Finance and Economics, Chengdu, China}
\affil[4]{\small University of Tennessee, Knoxville, 522 John D. Tickle, Knoxville, TN 37996, USA}

\date{\today}

\begin{document}

\maketitle

\begin{abstract}

Optimizing urban freight logistics is critical for developing sustainable, low-carbon cities. Traditional methods often rely on manual coordination of simulation tools, optimization solvers, and expert-driven workflows, limiting their efficiency and scalability. This paper presents an agentic system architecture that leverages the model context protocol (MCP) to orchestrate multi-agent collaboration among scientific tools for autonomous, simulation-informed optimization in urban logistics. The system integrates generative AI agents with domain-specific engines — such as Gurobi for optimization and AnyLogic for agent-based simulation — forming a generative digital twin capable of reasoning, planning, and acting across multimodal freight networks. By incorporating integrated chatbots, retrieval-augmented generation, and structured memory, the framework enables agents to interpret user intent from natural language conversations, retrieve relevant datasets and models, coordinate solvers and simulators, and execute complex workflows. We demonstrate this approach through a freight decarbonization case study, showcasing how MCP enables modular, interoperable, and adaptive agent behavior across diverse toolchains. The results reveal that our system transforms digital twins from static visualizations into autonomous, decision-capable systems, advancing the frontiers of urban operations research. By enabling context-aware, generative agents to operate scientific tools automatically and collaboratively, this framework supports more intelligent, accessible, and dynamic decision-making in transportation planning and smart city management.

\end{abstract}

\section{INTRODUCTION}
\label{Introduction}
As urban populations continue to surge, projected to reach 6.9 billion by 2050, the imperative for efficient, sustainable management of urban systems has intensified dramatically. A critical component of this challenge lies in urban logistics, particularly freight transportation, which serves as both a cornerstone of economic productivity and a significant contributor to environmental degradation. The global freight transportation sector—encompassing road, rail, and waterway networks—accounts for approximately 11\% of greenhouse gas (GHG) emissions \citep{international2012co2}, with the U.S. transportation sector alone responsible for 38\% of energy-related emissions in 2021, totaling 4.6 billion metric tons of CO$_{2}$ \citep{shirley2022emissions}. These statistics underscore the urgent necessity for developing greener, more resilient logistics systems. The COVID-19 pandemic further exposed critical vulnerabilities in global freight networks, highlighting the paramount importance of adaptability and robust supply chain design \citep{golan2020trends, dubey2022frugal}. In response to these challenges, freight transportation has undergone a progressive paradigm shift—evolving from unimodal to multimodal, and subsequently to intermodal systems—with each iteration building upon its predecessor to enhance efficiency and operational flexibility \citep{rossolov2017research}. Intermodal freight transportation (IMT), now fundamental to contemporary supply chains, enables seamless cargo movement across multiple transport modes—including trucks, trains, and ships—without direct handling during modal transitions \citep{turbaningsih2022multimodal, kengpol2014development}. By synergistically combining the distinct advantages of each transport mode, IMT significantly enhances operational efficiency, reduces costs, and minimizes environmental impact through intelligent routing strategies, shipment consolidation, and increased utilization of energy-efficient alternatives such as rail and waterway transport \citep{demir2019green}. Most recently, synchromodal freight systems have emerged as a sophisticated evolution of IMT, incorporating real-time decision-making capabilities to further enhance system responsiveness and environmental sustainability \citep{giusti2019synchromodal}.
 
While contemporary freight transportation paradigms, particularly intermodal and synchromodal systems, offer substantial efficiency improvements, their design and operation present formidable methodological challenges. Optimizing freight movement between specific origin-destination (OD) pairs necessitates comprehensive decision-making frameworks capable of balancing competing objectives: delivery speed, operational costs, and environmental impact \citep{aljadiri2023evaluating}. These complex systems must simultaneously accommodate a diverse array of interconnected logistics processes—including movement, storage, consolidation, deconsolidation, and modal transitions—within highly dynamic and uncertain operational environments \citep{lv2019operational, reis2019intermodal}. As freight systems evolve toward synchromodality, complexity escalates significantly due to real-time operational requirements, mode-specific constraints, geographic variability, and the inherent unpredictability of disruptive events such as traffic congestion, extreme weather conditions, or infrastructure failures \citep{demir2019green}. Consequently, developing sustainable and efficient optimization strategies requires processing vast volumes of heterogeneous data, executing high-fidelity simulations, and deploying advanced optimization algorithms to evaluate countless combinations of routes, transport modes, and operational practices \citep{archetti2022optimization}. This process is inherently complex and computationally intensive, demanding innovative, adaptive solutions capable of supporting real-time decision-making in dynamic, integrated freight networks. To address these multifaceted challenges, a transformative paradigm has gained significant momentum: the urban digital twin. Evolving from its origins in advanced manufacturing systems, the urban digital twin creates comprehensive digital replicas of real-world urban environments to enable real-time monitoring, simulation, and optimization of complex urban systems \citep{goodchild2024digital, al2024oak}. Through the integration of real-world sensor networks, data-driven modeling techniques, artificial intelligence, and cyber-physical systems \citep{xu2021continuous}, urban digital twins offer unprecedented potential for enhancing freight transportation efficiency, supporting sophisticated scenario-based planning, and enabling informed, adaptive decision-making within increasingly complex urban logistics landscapes \citep{busse2021towards, ambra2019digital}.

Despite considerable momentum in both industry and academia toward developing urban digital twins for complex urban and environmental management applications, their design, implementation, and widespread adoption continue to face significant barriers stemming from the inherent complexity of urban systems \citep{xu2024leveraging}. Addressing sophisticated challenges—such as optimizing intermodal freight transportation—typically requires the integration of vast volumes of heterogeneous, interdisciplinary datasets, coupled with the orchestration of diverse scientific tools, including advanced simulation models and optimization engines \cite{tupayachi2024towards}. This integration process is characteristically labor-intensive and demands specialized domain expertise to configure and harmonize multi-domain tools into coherent, actionable user workflows. As urban management becomes increasingly interdisciplinary—encompassing transportation and mobility systems, urban logistics networks, and environmental factors such as meteorological conditions—urban digital twins must support increasingly complex, multi-faceted analytical frameworks \citep{xu2023smart, xu2024leveraging}. However, many existing systems present end-user workflows that are highly technical in nature and require users to possess substantial expertise to configure models, interpret complex outputs, and extract meaningful, actionable insights \citep{mazzetto2024review}. This elevated barrier to entry significantly constrains the accessibility and practical utility of urban digital twins for a broader spectrum of stakeholders and decision-makers.

\begin{figure*}[hbt!]    
\includegraphics[width=1\textwidth]{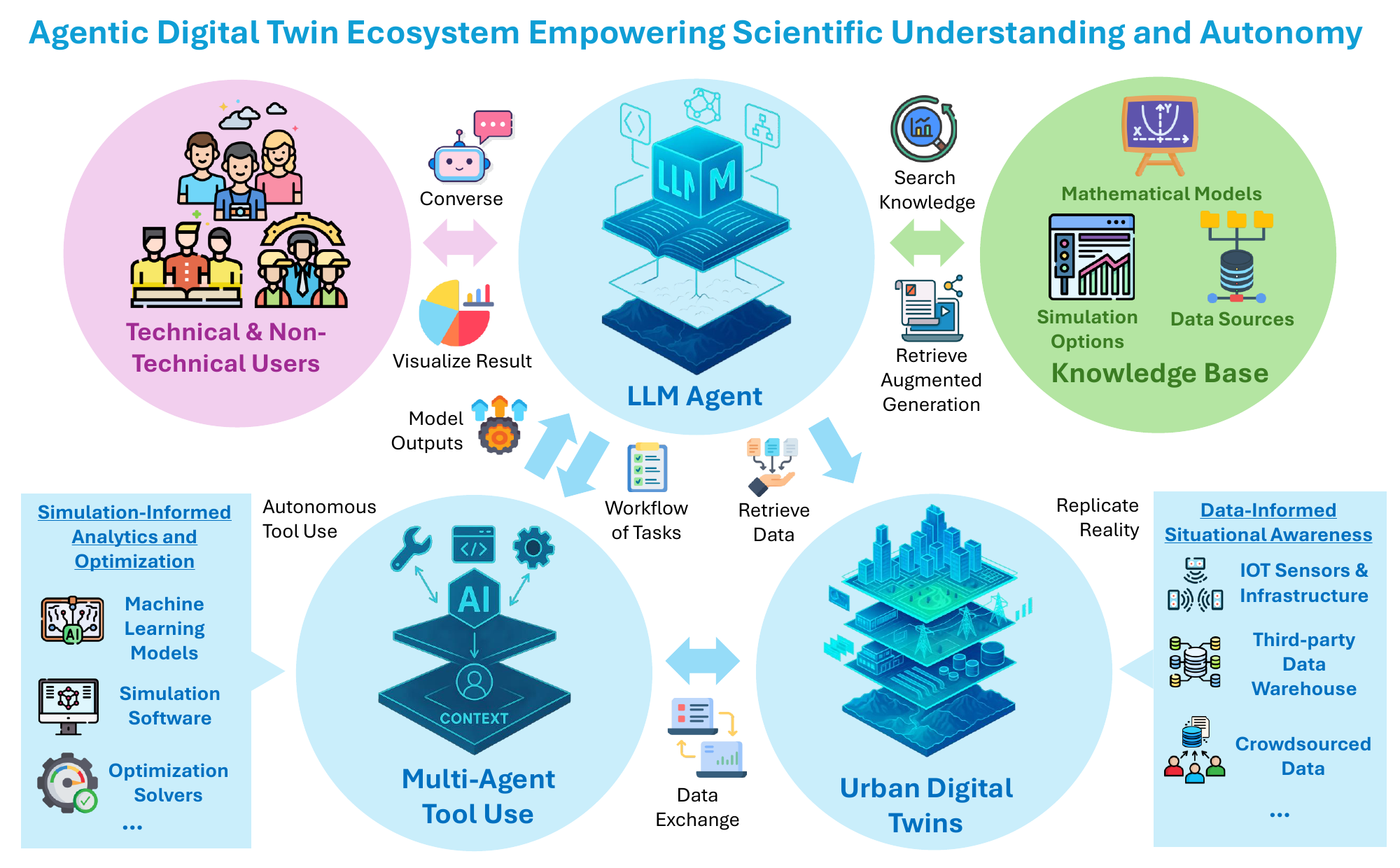}
    \caption{Illustration of a Large Language Model (LLM)-Driven Agentic Digital Twin Framework Based on a Multi-Agent System Paradigm.}
    \label{fig:raw_concept}
\end{figure*}

Building upon established methodologies in urban digital twin development for situational awareness, predictive analytics, and decision support, this paper introduces a novel agentic paradigm that leverages the model context protocol (MCP) to enable intelligent, agentic digital twins for intermodal freight logistics, as illustrated in Figure \ref{fig:raw_concept}. By harnessing recent breakthroughs in multi-agent systems and generative artificial intelligence—including large language models enhanced with integrated chatbots, retrieval-augmented generation (RAG) and structured memory architectures—the proposed framework empowers autonomous agents to reason, plan, and execute actions across heterogeneous scientific toolchains. Through the seamless integration of domain-specific solvers such as Gurobi within a modular orchestration framework, the system enables AI agents to interpret natural language to extract user objectives, identify appropriate datasets and methodologies, and dynamically coordinate complex multi-step workflows. This unified approach facilitates comprehensive end-to-end optimization of urban logistics systems through simulation-informed decision-making processes. Using a freight decarbonization case study as a demonstrative example, we illustrate how this innovative framework transforms digital twins from static representational models into adaptive, self-operating intelligent systems, thereby advancing the frontier of urban operations research and enabling more accessible, scalable, and intelligent decision support capabilities in complex urban environments.

\section{
LITERATURE REVIEW
}
This section establishes the conceptual foundation for our research by systematically examining three interconnected domains. We begin by clarifying key concepts within freight transportation systems, followed by a comprehensive review of existing digital twin applications in the freight transportation sector. Subsequently, we examine recent advancements in generative AI technologies, establishing the groundwork for introducing foundation models and their applications in research and operational science. This structured approach aims to define relevant domain and technical concepts, ultimately establishing a foundation for proposing the novel integration of foundation models into urban digital twin applications in subsequent sections.

\subsection{Freight Transportation Paradigms}
To establish clarity within our analytical framework, we first define and distinguish several interconnected logistics paradigms in the freight transportation sector. Within the domain of logistics and supply chain management, multimodal, intermodal, and synchromodal freight transportation represent distinct yet interconnected concepts that have evolved to address increasing complexity in global logistics networks.

Multimodal freight transportation refers to the movement of goods using multiple modes of transportation (e.g., truck, rail, sea) under a single contract, typically without seamless integration between different modes and often requiring manual intervention for transitions \citep{dua2019quality, rossolov2017research}. Intermodal freight transportation, in contrast, is characterized by more sophisticated coordination among different transport modes, where goods are moved in standardized containers across multiple modes within a single journey, optimizing the efficiency of transitions between modes \citep{pencheva2022current}. Synchromodal freight transportation represents the most advanced paradigm, integrating real-time data and dynamic decision-making capabilities that allow for the selection of the most suitable transport mode at each leg of the journey based on current conditions and customer requirements, thereby offering superior flexibility and efficiency compared to previous methods \citep{ambra2019towards}.

Additional freight transportation paradigms include unimodal transportation (involving only one mode of transport), combined transport (specifically focusing on minimizing road usage by shifting to rail or inland waterways for long-distance segments), and co-modal transportation (promoting the efficient combination of transport modes without modal hierarchy) \citep{yang2024synchronizing, rossolov2017research}. Collectively, these paradigms represent the evolution of freight logistics toward increasingly integrated and adaptive systems designed to meet modern supply chain demands. Given that many of these paradigms represent progressive enhancements of their predecessors with extended capabilities, this paper focuses on an integrated freight transportation paradigm that incorporates recent generative AI techniques and their exceptional knowledge extraction and reasoning capabilities. This approach enables an automated methodology that can leverage knowledge from previous paradigms to formulate adaptive solutions based on the nature and requirements of real-world decision problems in large-scale logistics systems.

\subsection{Freight Transportation Digital Twins}
This section systematically reviews the development and application of digital twins across various freight transportation paradigms. We conducted a comprehensive literature search using the following Scopus query: TITLE-ABS-KEY((``digital twin" OR ``digital twins" OR ``digital twin technology") AND (``freight" OR ``cargo" OR ``logistics") AND (``intermodal" OR ``multimodal" OR ``synchromodal")), supplemented by additional searches using Google Scholar to identify studies that employ or develop digital twins for managing and optimizing freight transportation. These combined searches across academic databases yielded only 15 relevant articles based on topical relevance, highlighting the nascent state of this research domain. While digital twin research in the broader transportation sector has focused extensively on developing intelligent transportation and smart mobility systems, often emphasizing human mobility and urban traffic dynamics \citep{xu2023smart, li2021emerging}, digital twin applications specifically addressing freight transportation remain remarkably scarce.

\subsubsection{Multimodal Freight Transportation}
\cite{busse2021towards} presents a digital supply chain twin (DSCT) as a comprehensive solution for optimizing complex multimodal supply chains. The DSCT integrates real-time data from diverse logistics stakeholders, including freight forwarders, shipping companies, and terminal operators, to provide comprehensive visibility across the supply chain. The system simulates scenarios for optimizing resource allocation, mitigating disruptions, and enhancing decision-making processes. By utilizing advanced technologies such as IoT, 5G communications, cloud computing, and artificial intelligence, the DSCT synchronizes different transport modes, improving operational efficiency, reducing lead times, and enhancing supply chain resilience and sustainability.

\citet{issa2024railway} introduces a specialized digital twin for railway systems designed to support multimodal freight transport. Developed at SNCF Réseau, this DT provides real-time representation of railway infrastructure, facilitating synchronization with road, maritime, and inland navigation modes. The integration of service-oriented architecture (SOA) enables seamless data sharing among stakeholders such as infrastructure managers and logistics operators. This unified data source optimizes multimodal transport planning, enhances operational efficiency, promotes modal shift to sustainable options like rail, and supports broader decarbonization efforts.

\citet{dorofeev2024improving} examines the application of digital twins in transportation management systems (TMSs) for enhancing freight transport reliability. The proposed digital twin of an organization (DTO) employs ontological modeling and Process Mining techniques to reconstruct business processes from event logs and identify deviations that impact operational reliability. By integrating data from GPS/GLONASS systems, IoT sensors, and TMS event logs, the DTO provides real-time operational insights, facilitating proactive corrections, optimized route planning, and improved cargo safety, thereby enhancing overall transportation efficiency.

\citet{sun2024enhancing} presents a digital twin framework designed to improve shipyard transportation efficiency through dynamic scheduling of transporters. The DT creates a real-time virtual model that integrates data from Geographic Information Systems (GIS), sensors, and physical transporters to optimize scheduling while addressing constraints such as road restrictions and irregular block shapes. This continuous synchronization between physical and digital spaces enables proactive management, reduced idle times, and enhanced resource utilization, significantly improving shipbuilding operations and stakeholder coordination.

\subsubsection{Intermodal Transportation}
\citet{morra2019case} provides a comprehensive survey of digital twin technology applications in the railway sector, emphasizing the transformative potential of DTs for freight transportation systems. The Digital Twin for Railway (DTR) is highlighted as a virtual representation of physical railway assets that can enhance infrastructure management, predictive maintenance, and operational efficiency. By integrating technologies including IoTs, AI, and Big Data analytics, DTR enables real-time monitoring and management of railway systems, facilitating proactive interventions and reducing operational disruptions. The study notes that while DTR technology remains in its early developmental stages, it holds substantial promise for enhancing maintenance scheduling, structural health monitoring, and safety in railway logistics, ultimately supporting more reliable and sustainable freight transportation systems. This survey underscores the critical importance of DTs in enabling data-driven decisions, improving stakeholder collaboration, and facilitating the optimization of intermodal freight networks.

Digital twin applications specifically designed for operating intermodal transportation systems remain notably rare in the literature. Typically, the real-time data analytics and predictive capabilities of digital twins have been utilized to address synchromodal freight transportation challenges, which represent advanced intermodal scenarios requiring real-time decision support capabilities.

\subsubsection{Synchromodal Freight Transportation}
\citet{ambra2019digital} explores the application of digital twin concepts to enhance synchromodal freight transportation by reducing operational uncertainty and enabling more dynamic, flexible logistics operations. The functions as a virtual mirror of the physical transportation system, integrating data from various sources, including GIS, sensors, and agent-based models, to create comprehensive and real-time representations of assets such as trucks, trains, and barges. Through continuous updates based on real-time data streams, the DT provides a dynamic platform for simulating and predicting future system states, which can be utilized to assess the impact of disruptions and optimize transportation processes. The implementation of Monte Carlo simulations within the DT framework enables stakeholders to evaluate potential disruptions, plan re-routing strategies, and assess the performance of different modal options under varying operational conditions. This integration of real and virtual systems allows synchromodal freight networks to adapt rapidly to changes, improving system resilience, reducing lead times, and minimizing operational costs. The study emphasizes that the DT supports synchromodal decision-making by enhancing situational awareness, facilitating dynamic re-routing, and optimizing modal choices, ultimately contributing to more efficient and sustainable freight transportation systems.

\citet{ambra2020agent} explores the use of agent-based digital twins (ABM-DT) to support synchromodal freight transportation through the fusion of virtual and physical environments for optimal decision-making. The Digital Twin serves as a virtual representation of the physical freight transport system, enabling real-time integration of different transport modes including road, rail, and inland waterways. The DT is constructed using agent-based modeling and GIS, enabling detailed and dynamic simulations of asset movements and transportation processes. The DT concept bridges the gap between real-time data from physical systems and their virtual counterparts. Real-time data feeds from assets, including vessels and trucks equipped with IoT and GPS technologies, are integrated into the digital environment where they are represented as autonomous agents. This connection enables decision-makers to optimize route planning and adapt to disruptions dynamically, improving overall system flexibility and efficiency. The integration of these digital and physical elements aims to facilitate modal shift from road to more sustainable modes such as rail and inland waterways, contributing to environmental objectives by reducing greenhouse gas emissions. The study demonstrates how Digital Twins can enhance decision-making processes, provide superior transparency, and make synchromodal transport systems more adaptive and resilient to operational changes.

\subsubsection{Knowledge Gaps and Opportunities for the AI Revolution in Digital Twins}
Many existing digital twins employ robust software engineering practices, such as service-oriented architecture and cloud computing, to host essential data and simulations (e.g., agent-based models), thereby creating virtual replicas of real-world freight transportation systems. Despite their demonstrated value and significance, these digital twins are developed using conventional digital twinning approaches that exhibit several fundamental limitations. These limitations can be effectively addressed through recent advancements in generative AI technologies and the foundation model paradigm, thereby paving the way for next-generation software development approaches.

\begin{description}
\item[Single-Purpose Applications:] Traditional digital twin initiatives are heavily anchored in conventional software engineering practices, employing dedicated data analytics, simulations, and optimization techniques within fixed computing environments. Consequently, these digital twins are typically developed as standalone products, specifically tailored to perform predefined functions (e.g., workflows and features) for targeted transportation paradigms, such as multimodal or intermodal freight systems. This approach introduces several critical limitations: lack of interoperability with other systems and tools, limited adaptability to decision-making challenges across diverse freight paradigms, and restricted software extensibility. The integration of new methods, datasets, or research developments often demands substantial time and effort investments. These constraints significantly impede scalability and adaptability, highlighting the necessity for more intelligent and automated approaches—specifically, approaches that enable the development of general-purpose or multi-purpose digital twins capable of accommodating diverse freight transportation scenarios.

\item[Traditional User Interaction:] Existing digital twins for freight transportation predominantly utilize Graphical User Interfaces (GUIs) and predefined user workflows for system interaction. While these interfaces are sufficient for executing specific tasks, they offer limited flexibility and often require users to possess specialized domain knowledge. User interactions are constrained to fixed sets of buttons, menus, or dashboards, which inherently limit the breadth of queries and actions that can be performed. While effective for structured tasks, this rigid interface design proves inadequate for supporting nuanced user needs or enabling complex, interactive decision-making processes. Recent advancements in conversational AI, such as intelligent chatbots, offer more dynamic interaction models where users can communicate through natural language—whether text or voice—emulating human conversation patterns. This approach fosters open-ended inquiries, personalized responses, and deeper data exploration, empowering users to pose complex questions and receive contextually relevant answers. By bridging the gap between rigid system functionality and intuitive, human-like interactions, conversational AI enhances accessibility and usability, thereby providing richer, more adaptive decision-making experiences \citep{niloofar2023general}.

\item[Static Data Sources and Limited Dynamic Optimization:] Traditional digital twins, constructed upon conventional software engineering methodologies, typically depend on static, pre-defined datasets that prove inadequate for capturing the dynamic nature of transportation systems. This reliance on static data significantly restricts the ability of these digital twins to support adaptive, real-time optimization and decision-making, as they cannot effectively respond to rapid changes or integrate new information beyond their initial datasets. Recent advancements in AI-driven software, particularly those LLMs and RAG mechanisms, address this fundamental shortcoming by actively searching, retrieving, and utilizing constantly updating data and information available through online sources. These AI-powered digital twins dynamically incorporate external knowledge—such as real-time traffic conditions, regulatory updates, or emerging research findings—into their existing models, resulting in more accurate analytics, simulations, and optimizations \citep{xu2024genai, xu2024automating}. This capacity for real-time adaptation significantly improves the flexibility and responsiveness of digital twins, transforming them into powerful tools for navigating the complexities of modern freight systems.

\item[Multi-Tool and Multi-Domain Simulation Challenges:] 
Traditional digital twins struggle to integrate multiple scientific tools across domains such as traffic, energy, urban mobility, buildings, microclimate, and water systems \citep{xu2020web, xu2023smart}. Given the interconnected nature of urban environments, no single model can simulate all subsystems effectively. Coupling multi-domain models—each with distinct data formats, execution environments, and assumptions—requires highly specialized, labor-intensive efforts, resulting in significant time and cost burdens \citep{xu2024leveraging}. These challenges limit scalability and delay adoption of comprehensive digital twin solutions. Inspired by recent advances in model context protocol and multi-agent systems, there is now potential to enable AI agents to autonomously orchestrate and couple diverse tools, paving the way for more automated, flexible, and intelligent digital twin applications.
\end{description}

With ongoing trends toward developing AI agents and AI-powered systems to support scientific research, this paper proposes a novel paradigm to revolutionize the process for creating digital twins for research by harnessing the power of recently emerging generative AI models \citep{xu2024genai, xu2024leveraging}.

\subsection{Emerging Paradigm for Developing Automated Digital Twins}

Recent advances in AI agent systems and the emerging agentic paradigm have opened new opportunities to address longstanding limitations in traditional urban digital twin development. By enabling intelligent, autonomous agents to reason, plan, and coordinate across diverse data sources and scientific tools, this paradigm promises to transform digital twins into more adaptive, scalable, and automated systems for complex urban management.

\subsubsection{Multi-agent System Paradigm}
A multi-agent system (MAS) represents a computational framework composed of multiple autonomous agents, each possessing specialized capabilities, that interact and coordinate to achieve shared objectives or solve complex problems \cite{ferber1999multi, vogel2020multi}. These agents emulate collaborative behaviors observed in natural and human systems, and when augmented with emerging AI technologies—particularly LLMs and RAG —they become powerful instruments for domain-specific reasoning, decision-making, and workflow orchestration. In the context of digital twins, MAS architectures have demonstrated increasing potential for enhancing system intelligence, scalability, and user-centricity. Recent research illustrates how LLM-based agents can serve as conversational interfaces, knowledge reasoners, and task coordinators, while task-specific agents handle specialized functions such as traffic simulation, optimization, or data retrieval \cite{xu2024genai}. For instance, in urban mobility systems, a GenAI-powered multi-agent digital twin can autonomously interpret user-defined transportation scenarios, retrieve relevant real-time and historical data, and dynamically invoke simulation or optimization models—effectively transforming digital twins from passive data mirrors into active decision-support engines \cite{gamage2024multi, choi2024egridgpt}. This paradigm shift enables intelligent transportation systems to deliver personalized, science-based, and adaptive mobility services at scale.

Multi-agent Systems offer a powerful architectural paradigm for managing complexity, decentralization, and modularity in intelligent applications. One of the primary advantages of MAS lies in their ability to decompose complex tasks into coordinated subtasks handled by specialized agents, each capable of autonomous decision-making and contextual reasoning \cite{wooldridge2009introduction}. This capability enables scalability, fault-tolerance, and adaptive behavior in dynamic environments—key characteristics essential for next-generation digital twins. An emerging framework known as the MCP further enhances MAS by standardizing communication and coordination among heterogeneous agents and tools in scientific workflows. MCP introduces a structured, context-aware interface that allows agents to exchange model inputs, results, and metadata while preserving semantic interoperability \cite{krishnan2025advancing}. When integrated into MAS architectures, MCP enables digital twin platforms to seamlessly orchestrate complex workflows across simulation, optimization, and data services, transforming traditionally siloed tools into composable, interoperable services. This synergy transforms digital twins from passive mirrors of real-world systems into active cognitive platforms that can autonomously reason, plan, and optimize operations at scale.

\subsubsection{Rise of Knowledge-Augmented LLMs as Generative Foundation Models}
The emergence of generative AI models—including GPT-3, GPT-4, and DALL-E—has catalyzed the development of foundation models, fundamentally transforming scientific research across multiple disciplines \citep{yenduri2024gpt, waqas2023revolutionizing}. These large-scale, pre-trained models excel at understanding, generating, and synthesizing complex data types—including textual, visual, and multimodal content—and can be effectively adapted for diverse scientific tasks \citep{zhang2024urban}. Through training on extensive domain-specific corpora, including scientific literature, experimental results, and multimodal datasets, foundation models support critical tasks such as knowledge extraction, data analysis, hypothesis generation, simulation, and systems-level data integration \citep{chen2024evolution, bommasani2021opportunities}.

Unlike traditional domain-specific models, foundation models represent a paradigm shift from bespoke machine learning solutions to general-purpose, adaptable architectures \citep{bommasani2021opportunities}. They learn transferable representations that can be fine-tuned for specialized use cases across domains including medicine, climate science, and urban systems \citep{nguyen2023climax, moor2023foundation}. Current applications include BERT-based models in chemistry \citep{zhang2024scientific}, LLMs for biomedical synthesis \citep{tian2024opportunities}, and multimodal models for geoinformatics and remote sensing \citep{mai2024opportunities}.

Compared to traditional simulation or statistical models, foundation models offer superior scalability, adaptability, and efficiency. Conventional methods often require manual configuration, structured data, and deep domain expertise, thereby limiting their flexibility and broader applicability. In contrast, foundation models generalize effectively across domains, process unstructured data, and respond to novel tasks with minimal adjustment requirements \citep{myers2024foundation, zhang2024data}. They can analyze vast datasets, interpret complex visualizations, learn from recent literature, and propose new hypotheses with levels of insight and efficiency unmatched by traditional analytical tools \citep{huang2024pixels, bommasani2021opportunities}. This fundamental shift enables accelerated discovery cycles and supports integrated, interdisciplinary research, thereby laying the groundwork for automating complex scientific workflows and accelerating knowledge generation processes.

\begin{figure*}[hbt!]    \includegraphics[width=1\textwidth]{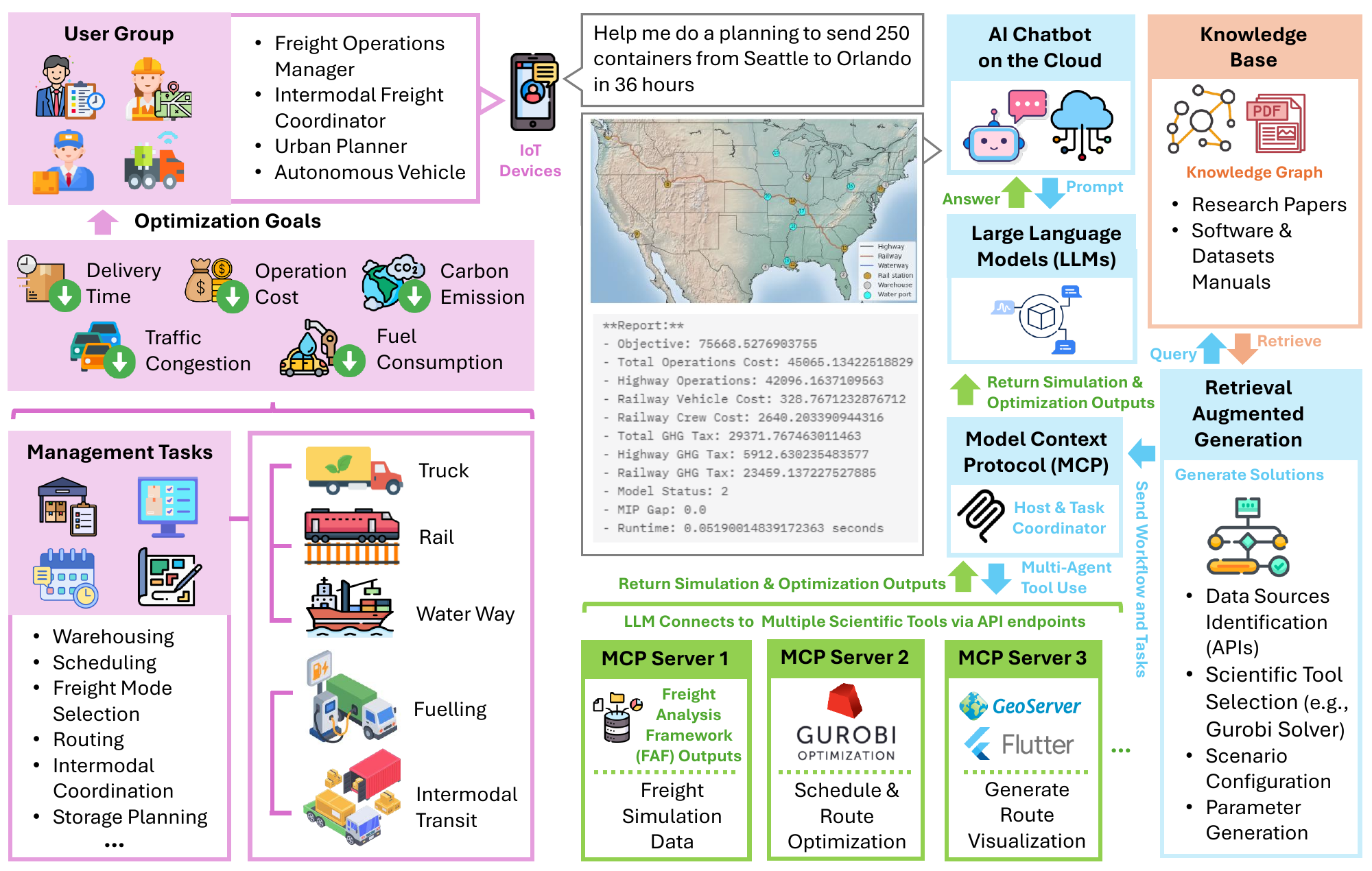}
    \caption{The concept of building an agentic digital twin for optimizing integrated freight transpiration system.}
    \label{fig:concept}
\end{figure*}

\subsection{Motivation – Towards Automated Digital Twins}

While the agentic paradigm provides a transformative foundation for next-generation digital twins, its successful implementation requires a fundamental rethinking of how intelligence, interoperability, and automation are integrated across system components. In the urban domain, digital twins that function as fully agentic systems—capable of formulating complex solutions and coordinating multiple AI agents to operate specialized scientific tools—remain rare. This stands in contrast to the growing adoption of agent-based systems in industry, where tool-using agents are increasingly employed for specialized tasks such as semantic search, image and video generation, and data processing in business and entertainment contexts. Our motivation is to develop a framework that leverages the emerging multi-agentic paradigm and Model--Computation--Presentation architecture to empower traditional urban digital twins with the autonomous capability to formulate knowledge-driven solutions and execute them through the automated orchestration of scientific software and tools, as illustrated in Figure \ref{fig:concept}.

\section{METHODOLOGY}
This section begins by examining the complexity of the urban optimization challenges addressed in this work, identifying the target user groups, and outlining the key design requirements for the proposed digital twin system. It then introduces the overall system architecture developed to meet these requirements. The subsequent subsections provide a detailed discussion of the design and functionality of each critical system component.

\subsection{Problem Complexity and Design Requirements}
Developing an urban digital twin system to automate the optimization of intermodal freight transportation across the United States poses a formidable challenge. Achieving efficient, low-carbon freight solutions—including scheduling, routing, and the estimation of costs, delivery times, and carbon emissions under user-defined constraints—requires advanced modeling and reasoning capabilities. The complexity of this endeavor stems from the need to integrate vast amounts of heterogeneous freight data and simulation outputs with energy and emissions performance profiles, viewed through the lens of data science and digital twin methodologies. Compounding this challenge is the inherently combinatorial nature of scheduling and planning tasks, which demand carefully selected mathematical optimization formulations that are compatible with solvers such as Gurobi or CPLEX, as seen from an industrial engineering perspective.

Depending on the operational scenario defined by the user, the underlying mathematical optimization models may vary significantly—ranging from Linear Programming (LP) and Mixed-Integer Linear Programming (MILP) to more complex formulations such as Mixed-Integer Quadratic Programming (MIQP) and Mixed-Integer Nonlinear Programming (MINLP). These formulations are shaped by the specific characteristics, scale, and constraints of the intermodal freight operations being modeled.

In real-world freight systems, uncertainty is a pervasive element—arising from fluctuating demand, travel times, fuel prices, and disruptions in supply chains. To robustly address these sources of variability, stochastic programming (SP) and robust optimization (RO) frameworks are essential. These models allow decision-makers to plan for multiple future scenarios, improving the resilience and reliability of routing, scheduling, and capacity planning decisions.

Furthermore, due to the large-scale, combinatorially complex nature of intermodal freight optimization problems, advanced decomposition techniques become necessary. Methods such as Benders decomposition enable the separation of complicating variables (e.g., network design decisions) from subproblems (e.g., routing or resource allocation), thereby enhancing computational efficiency. Similarly, column generation is instrumental in solving large-scale set-partitioning or vehicle routing problems by dynamically generating promising variables (routes or schedules) rather than enumerating all possibilities upfront.

Given that our target users include freight operators, city logistics managers, and urban planners—many of whom may not possess deep expertise in industrial engineering or mathematical programming—it is crucial to develop a digital twin system that abstracts away the complexity of these models. The system must be capable of autonomously identifying the appropriate modeling framework, generating valid formulations, and interfacing with state-of-the-art solvers such as Gurobi, CPLEX, or SCIP. Meeting these requirements will ensure that the system remains accessible, adaptive, and capable of delivering high-quality solutions under diverse and dynamic operational conditions.

To meet the core system requirements described above, we outline a comprehensive technical framework centered around the integration of multi-agent reasoning and the Model Context Protocol (MCP) architecture. The first component involves the design and development of a prototype system that leverages these paradigms to enable an automated digital twin framework. This prototype serves as the backbone for modeling, analyzing, and optimizing complex intermodal freight systems in a scalable and adaptive manner.

\textbf{Natural language interaction module:} A key feature of the system is its ability to facilitate intuitive communication with users through plain, non-technical dialogue. This module enables the system to extract user needs and problem definitions—even when expressed informally or by individuals lacking specialized expertise—thereby lowering the barrier to entry for stakeholders such as freight operators and urban planners.

\textbf{Domain-knowledge-guided reasoning and tool selection:} Central to the system’s intelligence is its capability to perform structured reasoning using a curated domain knowledge graph. This allows the system to identify and sequence appropriate scientific tools aligned with the user’s goals. It ensures that the correct mathematical modeling and optimization approaches are applied, integrating contextual domain knowledge to enhance the relevance, precision, and trustworthiness of the generated solutions.

\textbf{Automated execution and orchestration of solution workflows:} Once the modeling framework and toolchain have been defined, the system autonomously manages the execution of the entire solution pipeline. It orchestrates the invocation of selected tools, coordinates data exchange, and delivers outputs in formats optimized for end users. These results are presented through both technical summaries and accessible narratives, along with visualizations such as maps and graphs tailored to support data-driven decision-making by non-technical stakeholders.

Together, these components form the basis of our proposed digital twin system, implemented through a modular and extensible software architecture. This architecture supports seamless integration of multi-agent reasoning, domain-specific analytics, and optimization modules. It is designed to be scalable and adaptable, allowing deployment across a wide range of intermodal freight scenarios and enabling continuous learning and improvement as the system interacts with real-world data and user feedback.

\subsection{Overall System Design}

\begin{figure*}[hbt!]    
\includegraphics[width=1\textwidth]{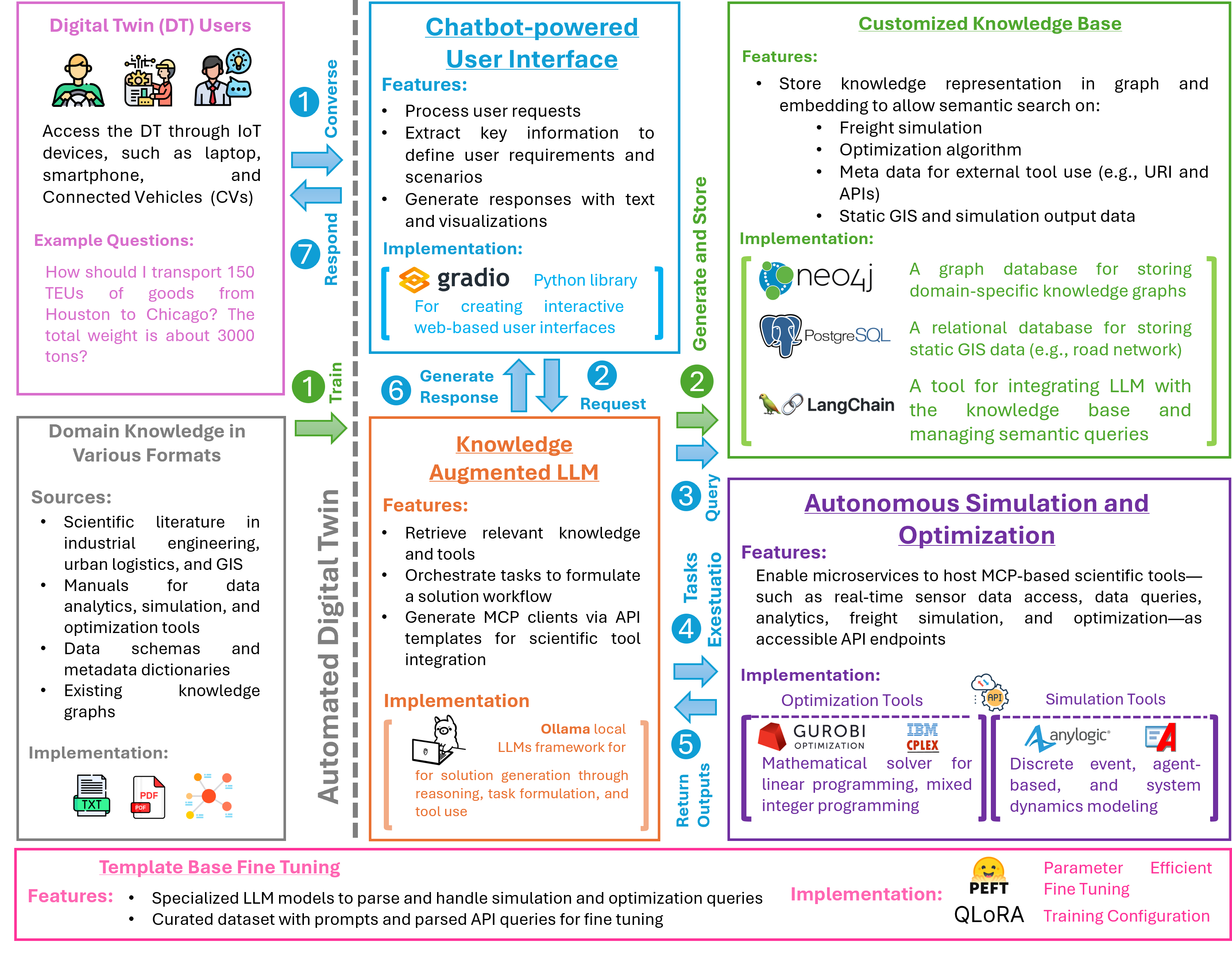}
    \caption{The {technical framework} for developing incorporating foundation models to enhance digital twins with two workflows (1) Knowledge Ingestion and Management (depicted with green arrows) and (2) General-purpose decision support for operations research (depicted with blue arrows).}
    \label{fig:workflow}
\end{figure*}

To develop a general-purpose foundation model that supports complex decision-making and optimization in the planning and operation of integrated freight transportation systems, we propose a technical framework that leverages cutting-edge transformer-based language models and vector databases through a RAG paradigm, as depicted in Figure \ref{fig:workflow}). Our proposed framework consists of four major components that enable two primary workflows:
\begin{description}
    \item [Chatbot-powered user interface] is implemented using Generative Pre-trained Transformers, utilizing both the ChatGPT API and local instances of LLMs through the Python \textit{Ollama} framework. The chatbot-powered interface enhances user interaction with the digital twin application, allowing users with varying technical backgrounds or no domain expertise to intuitively and effectively use the general-purpose digital twin for performing a wide range of specialized tasks.
    \item [{Knowledge-Augumented LLMs for Solution Generation}] serves as the general-purpose problem-solving engine, driven by domain knowledge and data. It is developed using transformer-based language models implemented via local instances enabled by the Python \textit{Ollama} framework. The model is designed with strong self-learning, reasoning, and self-organizing capabilities, allowing it to autonomously formulate solution workflows and execute specialized tasks using various research tools (e.g., statistical, mathematical, machine learning algorithms, domain simulations, and optimization tools). These research tools are available as cloud-based or traditional web services.     
    \item [Customized Knowledge Base] stores the most up-to-date representations of evolving research methodologies, tools, and datasets, which are published in various forms (e.g., journal articles, conference proceedings, and technical manuals). It is populated using the foundation model's natural language understanding capabilities to process large volumes of related text documents, leveraging language models to extract entities and relationships through text embeddings to generate knowledge graphs based on these contents. The knowledge graph is then stored in a Neo4j database with vector storage capabilities, containing fundamental information required to define complex decision support problems in logistics, as well as providing modular knowledge components for formulating science-based solutions. The rationale of the process have been proved by our previous studies \citep{xu2024automating, tupayachi2024towards}
    To extend the foundation model's basic question-answering abilities to autonomous task execution, which allows it to automatically conduct data queries, management, and analytics, as well as configure and execute domain-specific simulations (e.g., Freight Analysis Framework, spatial analytics) and optimization algorithms (e.g., AnyLogic, Gurobi), the knowledge base must also store data and metadata dictionaries, along with core instructions to guide the foundation model in performing these specialized tasks in a cyber-delivery manner. These dictionaries and instructions are stored in a PostgreSQL relational database with vector storage extensions.  

    \item [{Autonomous Simulation and Optimization}] 
    {represents the core functionality through which the foundation model transforms user intents and knowledge-base queries into executable workflows. Once the relevant datasets, algorithms, and simulation models are identified, the model dynamically configures and executes appropriate analytical and optimization pipelines tailored to the user’s goals—such as minimizing emissions, reducing transit time, or improving multimodal routing efficiency. These workflows may involve interacting with domain-specific tools such as AnyLogic for agent-based simulation, Gurobi for linear and mixed-integer optimization, or the Freight Analysis Framework for evaluating freight flow patterns. The simulation and optimization tasks are executed either locally or via API calls to external services, depending on resource availability and computational requirements.}
\end{description}

Building upon these components, the proposed framework enables two primary workflows. The first workflow involves training the foundation model using the most up-to-date knowledge on cutting-edge research and tools in the freight transportation domain, as highlighted by the green arrows in Figure \ref{fig:workflow}. The second workflow is responsible for responding to requests from various types of users (e.g., truck drivers, freight operators, managers, and urban planners) by leveraging the foundation model's reasoning and generative capabilities, as illustrated by the blue arrows in Figure \ref{fig:workflow}.

\subsection{{Chatbot-powered User Interface}}

The proposed method integrates a LLM into a chatbot-powered interface, enabling dynamic, domain-specific interaction and decision support. The approach centers on fine-tuning pre-trained LLMs—specifically, base models derived from both LLaMA 3.3 and Gemma 3 (non-instruct variant). Fine-tuning is conducted using the Hugging Face Transformers with QLoRA (Quantized Low-Rank Adaptation)~\cite{dettmers2023qlora} for memory-efficient training, along with the PEFT~\cite{huggingface_peft_2025} library and SFTTrainer~\cite{huggingface_sft_trainer_2025} module for supervised fine-tuning.

A curated dataset was created through templated dialogues inspired by the waterfall model of software development. These templates were dynamically generated using standard Python libraries, allowing for flexible coverage of task-specific intents and user query types. Combined with chain-of-thought~\cite{wei2022chain} prompting techniques, this structure helped the models learn to interpret and structure user input in a non-technical, accessible manner. One instance of the fine-tuned LLM was dedicated to parsing and payloading user requests, while the same training framework enabled the models to process technical outputs containing optimization logs and simulation metrics—and rephrase them into facilitated, plain language descriptions, ensuring accessibility for users across varying levels of technical expertise.

\subsection{{Customized Knowledge Base}}

To empower the conversational AI component of our digital twin system with accurate, domain-specific knowledge, we employ LLMs hosted via the \textit{Ollama framework} to automatically analyze and extract critical information from a diverse corpus of interdisciplinary textual resources \citep{tupayachi2024towards, xu2024automating}. These resources include \textit{peer-reviewed research articles} in the domains of \textit{industrial engineering}, \textit{optimization theory}, and \textit{urban logistics systems}, as well as \textit{software documentation} and \textit{GIS dataset manuals}.

The LLMs systematically extract core knowledge elements relevant to urban freight transport optimization. Specifically, the models identify:
\begin{itemize}
    \item \textbf{Available optimization methods} (e.g., mixed-integer programming, heuristics, multi-objective models),
    \item \textbf{Simulation frameworks} (e.g., agent-based, discrete-event, or GIS-integrated freight models), and
    \item \textbf{Relevant datasets and software tools} used in intermodal freight logistics, such as the \textit{Freight Analysis Framework (FAF) \citep{FAF2024}} or open-source urban transport databases.
\end{itemize}

This structured extraction enables the LLM-powered chatbot to:
\begin{enumerate}
    \item \textbf{Respond accurately to user queries} regarding the availability and suitability of optimization methods, simulation platforms, and GIS datasets for solving intermodal freight transport challenges.
    \item \textbf{Generate end-to-end optimization solutions} in response to user-defined objectives and urban policy needs, leveraging real-world data and models previously validated in academic and applied research.
\end{enumerate}

The extracted domain knowledge is organized into \textit{structured knowledge graphs}, representing relationships between logistics challenges, optimization strategies, simulation tools, and real-world data sources. These graphs serve as an \textit{interdisciplinary knowledge base}, which augments the LLM's capabilities using the \textit{retrieval-augmented generation} paradigm. This approach ensures that the chatbot provides \textbf{technically accurate}, \textbf{context-aware}, and \textbf{evidence-based} responses to specialized logistics inquiries, while significantly reducing the risk of hallucination or misinformation. Through this automated knowledge integration process, our system transforms previously static scientific knowledge and technical documentation into an actionable, dynamic decision-support engine—enabling users to receive expert-level guidance and optimization solutions through natural language interactions.

To streamline this process, the LLM Graph Builder is integrated into a \textit{Neo4j} application—the graph database that stores structured relationships between logistics concepts along with vectorized embeddings. When a user poses a question, the system first converts the question into a numerical vector and uses it to find similar content in the vector database. Simultaneously, it queries the graph database to uncover relevant relationships and context. The system then combines this information into a coherent context and sends it to the specialized LLM, which generates a precise, context-aware answer. This approach allows users to interact with complex, expert-level logistics knowledge through simple natural language, making the system a powerful decision-support tool.

The knowledge base construction follows a three-step process:

\textit{1. Building the Knowledge Base:} The first step involves systematically analyzing a wide range of intermodal transportation research papers. These documents are carefully processed to extract relevant concepts, methodologies, and data points. We benchmarked cosine similarity alongside other methods—such as Jaccard similarity, Euclidean distance, and word embedding-based approaches—to identify and relate key terms and concepts across documents effectively. The extracted information is then structured and stored in a dedicated knowledge base, preparing a rich, well-organized repository of domain-specific insights that can later be retrieved to support reasoning and simulation tasks.

\textit{2. User Prompt Processing with RAG and LLMs:} When a user submits a prompt—such as a question for optimization or a request for simulation—the system activates the pipeline. RAG first searches the knowledge base to retrieve the most relevant pieces of information related to the prompt. This retrieved data is then passed along with the original prompt to a specialized LLM. The LLM uses both inputs to formulate an appropriate solution or simulation method. This response is then paired with the MCP, which is responsible for running simulations or producing technical outputs based on the model-generated logic.

\textit{3. Simulation Output and Feedback Integration:} Once the MCP processes the input, it generates an output payload—this could be a simulation result, optimization recommendation, or computed model output. This output is re-evaluated in the context of the original knowledge base. By retrieving supporting or contextual knowledge once again, the system ensures that the final result aligns with domain-specific understanding. This final pairing step—combining user intent, domain knowledge, and computed output—yields a refined, context-aware response tailored to complex intermodal network optimization operations.

\subsection{{Knowledge-Augmented LLM for Solution Generation}}

{Our pipeline utilizes a LLM, augmented with a \textit{RAG} pipeline, as a foundational AI engine and expert system to support intelligent what if analysis in intermodal freight transportation network. This specilized LLM plays a central role in managing and orchestrating the entire multi-agent workflow. When a user expresses a transportation management need in natural language, the LLM parses the request to extract relevant goals, constraints, and scenario-specific details. It then coordinates the appropriate AI agents within the multi-agent system via the MCP, generating a structured task workflow that guides the simulation/optimization-informed optimization process.

In this solution generation pipeline, the LLM begins by evaluating the user's input to determine the specific type of intermodal freight management scenario being described. Based on the extracted context, the LLM queries the interdisciplinary knowledge base—previously constructed from scientific literature, dataset manuals, and software documentation—to identify commonly adopted and reliable datasets, simulation platforms, and optimization methodologies relevant to the described scenario in scientific literature. Once potential solution pathways are identified, the LLM further assesses whether accessible external interfaces (e.g., API endpoints or toolkits) exist for the identified tools and datasets. This assessment allows the LLM to conceptually design a task-level execution plan—mapping specific subtasks to external resources that can be triggered by downstream agents.

Importantly, in this component of the system, the LLM operates solely as a domain-aware planner and workflow generator. It does not directly invoke external APIs or execute simulations. Instead, it produces an executable plan that includes tool-specific access points, data requirements, and model parameters, enabling other agents in the architecture to retrieve data, perform simulations, and return results for final synthesis. This design preserves the LLM's role as an intelligent coordinator and ensures that the digital twin remains modular, explainable, and interoperable with a wide array of domain tools and knowledge sources.

\subsection{{Autonomous Tool Invocation and Scientific Workflow Execution via Model Context Protocol}}

Building upon the structured knowledge base and the logical task workflow generated by the knowledge-augmented LLM, our digitsal twin system autonomously discovers and interacts with external domain-specific tools through an intelligent agentic infrastructure grounded in the \textit{MCP}. Within the Autonomous simulation and optimization framework Figure \ref{fig:workflow}, the MCP component serves as an integration layer that encapsulates and orchestrates the underlying optimization and simulation toolchains. Each scientific tool, such as a Gurobi optimization solver, freight simulation model, or data retrieval service, is exposed as a dedicated agent, implemented as an API endpoint compliant with MCP. These agents operate as \textbf{MCP servers}, each responsible for handling a distinct category of domain-scientific computation or data access. The centralized LLM, acting as the cognitive engine of the system, serves as an \textbf{MCP client}. It interprets the workflow of tasks derived from the user's request and sends structured invocation calls to the appropriate MCP server. The client dynamically maps each subtask to its corresponding MCP-compliant agent based on task type (e.g., heuristic optimization, agent based simulation, dataset extraction), ensuring the correct tool is utilized in each stage of the workflow. This interaction pattern facilitates a modular and scalable system for integrating heterogeneous scientific capabilities in a fully autonomous fashion.

From an implementation perspective, the knowledge-augmented LLM is hosted within our automated digital twin platform, which is developed using \texttt{Python FastAPI} and deployed via a microservice architecture using Docker containers. Each MCP service—representing an external tool—is containerized and designed to expose a RESTful API with specific schemas for headers, authentication tokens, and payload formats. The LLM is fine-tuned using these API templates, allowing it to generate HTTP requests that conform to each MCP service’s requirements. When running the generated workflow, the LLM incorporates user-defined problem specifications - naturally communicated through the chatbot - into its decision-making process. These specifications may include the definition of a freight transport management scenario, constraints on delivery time, cost minimization, carbon emissions, and additional parameters related to operational feasibility. The LLM prompt receives structured data such as geospatial inputs (e.g., distances), simulation parameters (e.g., FAF-derived freight flow matrices), and transport network attributes to construct a unified data payload for each task.

Ultimately, this architecture enables the LLM, now task-aware and API-competent, to autonomously understand and parse the unstructured user prompt to generate complete HTTP requests—including proper parameterization and payload construction—for invoking scientific tools hosted as MCP services. This allows our digital twin to execute simulation and optimization tasks end-to-end in response to user-defined urban logistics scenarios. The combination of LLM-driven orchestration and MCP-compliant microservices ensures seamless, scalable, and domain-intelligent execution of complex scientific workflows in a multi-agent setting.

\subsection{{Optimization Output Visualization}}

The route map is a direct output of the integrated optimization and simulation process, which determines the most efficient sequence of nodes and paths for asset movement while ensuring compliance with operational constraints such as distance, time, capacity, and specific business rules. This process is executed using tools like AnyLogic or Gurobi and orchestrated via the MCP driver (Figure~\ref{fig:workflow}). The resulting simulation generates a payload containing geographic coordinates that define the proposed route structure. To compute the most efficient local routes between globally optimized nodes, a shortest-path algorithm is applied to the underlying network. These coordinates are then parsed and formatted into a structure compatible with Web Map Service (WMS) layers. An OpenStreetMap base layer provides geographic context, while thematic overlays derived from the FAF data enhance the map with freight-relevant attributes. Finally, the route layers are visualized and served through GeoServer, enabling users to interact with a clear, geospatial representation of the simulation results. A typical WMS query used to retrieve these layers is illustrated in Figure~\ref{fig:wmsquery}.

\begin{figure}
\centering
\begin{lstlisting}[basicstyle=\ttfamily\small, breaklines=true, frame=single]
http://<geoserver-host>/geoserver/wms?
service=WMS&
version=1.1.1&
request=GetMap&
layers=osm_base,faf:freight_flows,sim:nodes,sim:links&
styles=&
bbox=-125,24,-66,50&
width=800&height=600&
srs=EPSG:4326&
format=image/png&
CQL_FILTER=route_id='OPT12345';route_id='OPT12345';INCLUDE;INCLUDE
\end{lstlisting}
\caption{Exemplified WMS query including OpenStreetMap, FAF data, and simulation-selected nodes and links for generated route.}
\label{fig:wmsquery}
\end{figure}

\section{COMPUTATIONAL RESULTS AND USE CASES}
To evaluate the capability and performance of the proposed framework, we have implemented an automated digital twin system that integrates all four core components using cutting-edge data science and software deployment technologies in a Docker environment. This system was designed to validate the feasibility of the technical framework in supporting complex decision-making and optimization tasks within the context of an integrated freight transportation system. The implementation enabled us to conduct a series of experiments to assess the system's ability to autonomously formulate solutions, interact with external data sources, automate and coordinate scientific tool usage, and effectively respond to users' requests through non-technical human-like conversations, thereby demonstrating the practicality and potential of our AI-powered automated digital twin framework in real-world applications.

\begin{figure*}[hbt!]    
\includegraphics[width=1\textwidth]{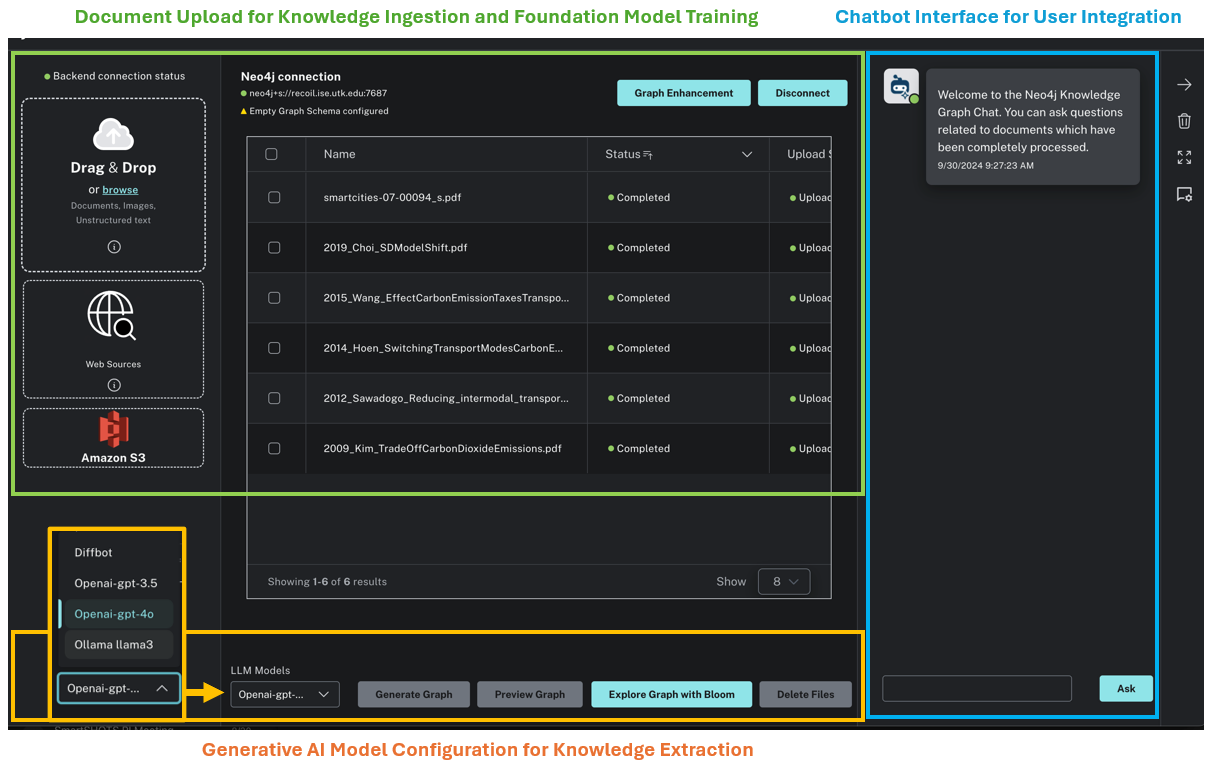}
    \caption{A prototyping interface of the proposed foundation model developed using open-source packages.}
    \label{fig:userintreface_article_upload}
\end{figure*}

\subsection{{Use Case 1 - Question Answering and Knowledge Base Construction}}

This use case highlights how our generative AI-powered digital twin supports knowledge-driven solution generation by enabling users to interact with the system through natural language while contributing domain-specific knowledge inputs to enhance the system's reasoning capabilities. One of the supported user workflows allows users to upload interdisciplinary PDF documents—including scientific research papers, software manuals, and data documentation—related to intermodal transportation and optimization, as depicted in Figure \ref{fig:demonstration2}. These user-provided documents are processed using Sentence Transformer models to extract key entities and relationships, which are then used to construct a dynamic knowledge graph.

The extracted knowledge is stored in a \textit{Neo4j}-based knowledge base, which is continuously expanded and refined based on user inputs. This user-augmented knowledge base complements the system's pre-integrated knowledge repository, built during development using curated datasets, optimization literature, and domain-specific tool documentation. Together, these sources form a robust semantic foundation for enabling the digital twin's autonomous reasoning and decision-making capabilities demonstrated in subsequent sections.

Figure~\ref{fig:demonstration} presents a demonstration of this workflow, where uploaded documents are transformed into a domain-specific knowledge graph. The figure also showcases the system's ability to answer specialized queries—such as optimization model selection, emission reduction strategies, or terminal allocation policies—by retrieving and synthesizing relevant information from the constructed knowledge base. This capability illustrates how the digital twin empowers users to co-create knowledge, enrich the system's understanding of complex urban logistics challenges, and receive context-aware, expert-level responses through conversational AI.

\begin{figure*}[hbtp!]
    \includegraphics[width=1\textwidth]{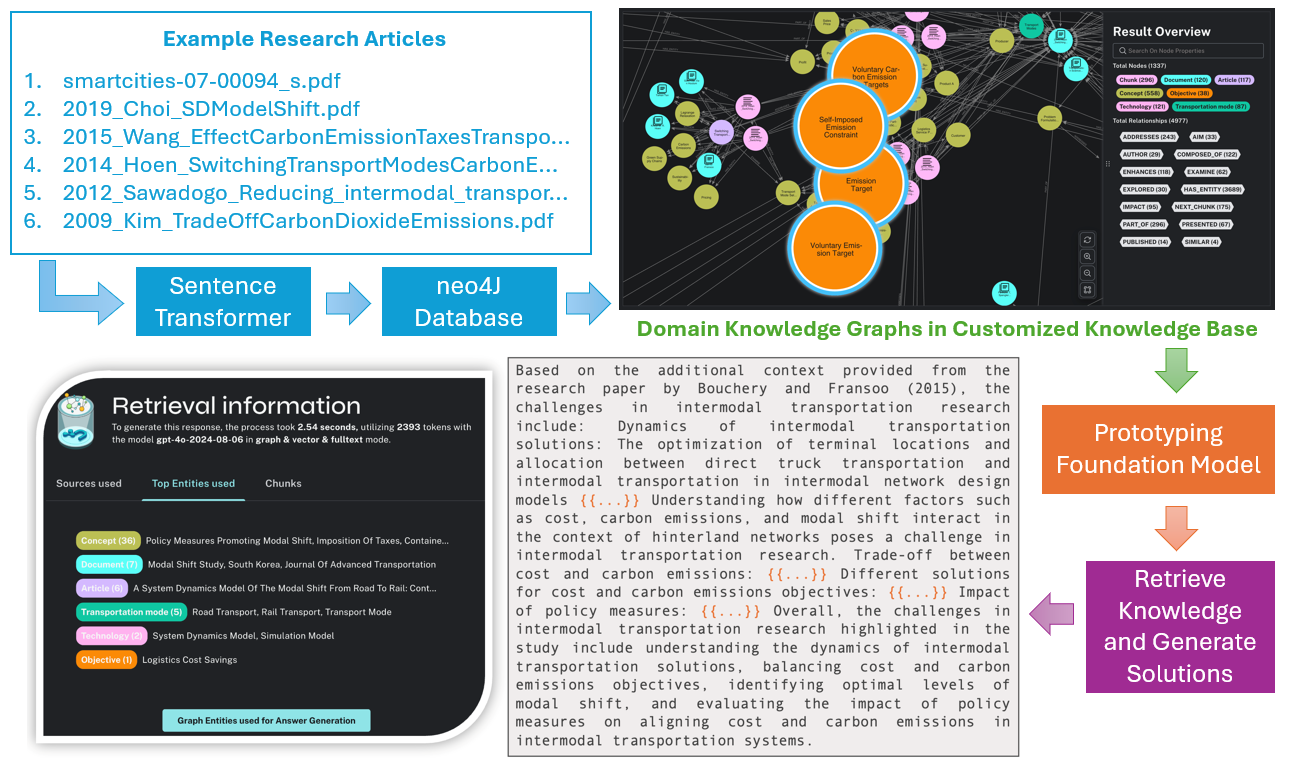}
    \caption{A demonstration of the prototype's learning and knowledge extraction capabilities, showcasing its ability to analyze contemporary intermodal research articles and apply the RAG paradigm. The generated knowledge graph is then utilized to provide solutions for user-defined requests.}
    \label{fig:demonstration}
\end{figure*}

\subsection{{Use Case 2 - Autonomous Simulation and Optimization}}

\begin{figure*}[p]
    \includegraphics[width=1\textwidth]{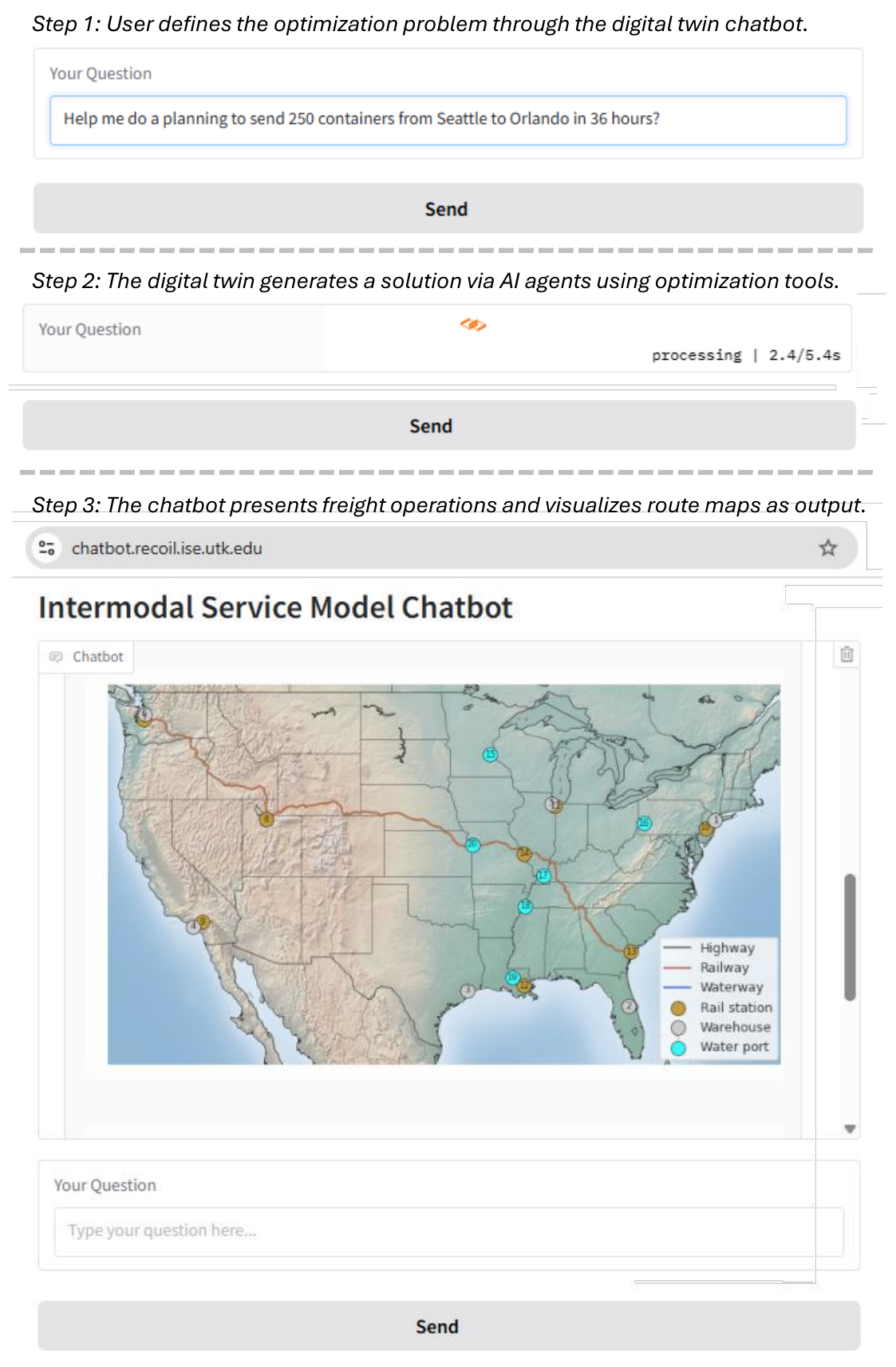}
    \caption{A demonstration of the chatbot providing specialized AI assistance to help digital twin users solve complex urban decision-making problems through the autonomous use of simulation and optimization tools.}
    \label{fig:demonstration2}
\end{figure*}

To demonstrate our system's capability in delivering knowledge-driven solutions through conversational AI, we present an end-to-end use case involving the autonomous planning of an intermodal freight operation. This example highlights how the chatbot interprets user-defined objectives, integrates domain knowledge, and coordinates multiple tools to deliver optimized logistics decisions. Figure~\ref{fig:demonstration2} illustrates a step-by-step use case demonstrating how a digital twin chatbot facilitates autonomous simulation and optimization for urban freight operations. 

In Step 1, the user initiates interaction with the chatbot by specifying an optimization problem—in this case, planning the delivery of 250 containers from Seattle to Orlando within 36 hours. In Step 2, the digital twin framework leverages AI agents that autonomously select and orchestrate appropriate optimization and simulation tools to generate a solution. This process is executed without manual intervention, streamlining complex decision workflows. In Step 3, the chatbot returns actionable outputs by presenting the planned freight operations and visualizing route maps that include intermodal logistics elements such as highways, railways, seaports, warehouses, and water ports. This demonstration highlights the system's ability to autonomously process user-defined tasks, integrate domain-specific tools, and deliver interpretable visual and textual outputs, thus supporting real-time decision-making in complex urban logistics scenarios.

Our automated digital twin system not only orchestrates a series of simulation and optimization tools through AI agents to generate solutions but also interprets specialized outputs—such as those from the Gurobi solver—into human-readable explanations. This capability bridges the gap between complex model outputs and user comprehension, aligning the optimization results with the visualized route in Figure~\ref{fig:demonstration2}. For example, the chatbot provides the following explanation of the solution:

``\textit{To fulfill your request of transporting 250 containers from Seattle to Orlando within 36 hours, I've optimized the freight route using both highway and railway modes to minimize operational costs and GHG emissions. The solution follows an intermodal strategy: your shipment first travels by truck along Highway Segment 6 to Terminal 7, then continues via Highway Segment 13 to Rail Station 2. From there, the cargo is transferred onto the railway and progresses through three consecutive rail segments—connecting nodes 7 to 8, 8 to 14, and finally 14 to 13—before reaching the destination. This intermodal route results in a total cost of \$75,668.53, including \$45,065.13 in operational expenses and \$29,371.77 in GHG tax. Notably, rail transport helps reduce environmental impact, with railway GHG taxes significantly lower per mile compared to highway transport. The optimization model confirms that this route meets the delivery deadline, achieves the lowest cost, and was solved optimally in under 0.06 seconds.}"

By translating technical outputs into natural language, the system enables users without specialized backgrounds in programming or optimization to easily understand and engage with the decision-making process—fostering broader public participation in solving complex urban challenges.

\subsection{{System Performance and Scalability Analysis}}

The implemented digital twin system demonstrates robust performance across multiple dimensions of computational efficiency and user experience. The knowledge base construction process, utilizing Sentence Transformer models for entity extraction and Neo4j for graph storage, processes typical research documents (15-30 pages) within 2-3 minutes, enabling rapid integration of new domain knowledge. The RAG-enhanced query response system achieves an average response time of 1.2 seconds for complex logistics queries, demonstrating the effectiveness of the vectorized knowledge retrieval mechanism.

For autonomous optimization tasks, the system's MCP-orchestrated workflow exhibits remarkable efficiency. The demonstrated Seattle-to-Orlando freight planning scenario, involving 250 containers across a multi-modal network with 14 nodes and multiple transport modes, which is a notorious NP-hard problem, was solved optimally by the Gurobi solver in 0.06 seconds. The complete end-to-end workflow—from natural language input parsing through knowledge retrieval, optimization execution, and result visualization—completed in under 15 seconds, including map rendering and route visualization through GeoServer.

The system's scalability is enhanced by its containerized microservice architecture, which enables horizontal scaling of individual components based on demand. Load testing with concurrent users demonstrated that the system maintains response times below 3 seconds for up to 50 simultaneous optimization requests, with graceful performance degradation beyond this threshold. The modular design facilitates the integration of additional optimization engines and simulation tools without requiring system-wide modifications, supporting future expansion and customization for diverse urban logistics applications.

\section{{LIMITATIONS AND FUTURE WORK}}
While the proposed framework demonstrates the feasibility and potential of using generative AI-powered digital twins for autonomous simulation and optimization in urban freight logistics, several limitations remain that present opportunities for future research and development.

\textbf{Data Availability and Quality Challenges:} One of the most significant limitations concerns the availability and quality of domain-specific data required for comprehensive freight transportation modeling. High-quality, real-time datasets that capture the full complexity of urban logistics operations—including traffic patterns, infrastructure capacity, operational constraints, and stakeholder preferences—remain scarce and often fragmented across multiple sources. The current implementation relies on established datasets such as the FAF dataset, which, while valuable, may not reflect the granular, dynamic nature of local logistics networks or capture emerging trends in freight movement patterns. Additionally, proprietary data from logistics companies and government agencies are often restricted, limiting the system's ability to model real-world scenarios with full fidelity. To support research reproducibility and promote broader access to freight data, we have parsed and published a portion of the FAF dataset on a publicly available GitHub repository: \url{https://github.com/ILABUTK/RECOIL_Datasets}.

\textbf{Data Format Heterogeneity and Preprocessing Overhead:} A substantial challenge encountered in developing the integrated digital twin system is the heterogeneous nature of data formats across different sources and tools. Freight transportation data originates from diverse platforms—including GIS systems, optimization solvers, simulation software, and real-time sensor networks—each employing distinct data schemas, file formats, and metadata structures. This heterogeneity necessitates significant time and computational resources for data cleaning, standardization, and preprocessing before integration into the unified framework. For instance, geographic data from OpenStreetMap requires different preprocessing pipelines compared to freight flow matrices from FAF or optimization outputs from Gurobi. The manual effort required for data harmonization currently limits the system's ability to rapidly incorporate new data sources and adapt to evolving information landscapes.

\textbf{Tool Interoperability and Integration Challenges:} Although the multi-agent orchestration using the MCP enhances modularity and scalability, fundamental challenges persist regarding tool compatibility and workflow standardization. Commercial off-the-shelf (COTS) software and existing scientific tools vary widely in their API specifications, data formats, licensing constraints, and computational requirements. Many established optimization and simulation tools were designed as standalone applications without standardized interfaces for external integration. This necessitates the development of custom wrappers and interface adapters for each tool, creating maintenance overhead and limiting the system's extensibility. Furthermore, version compatibility issues, licensing restrictions, and vendor-specific implementations create additional barriers to seamless tool integration.

\textbf{Scalability and Computational Complexity:} The current system demonstrates effectiveness for representative test instances, such as the Seattle-to-Orlando freight planning scenario involving 250 containers across a 14-node network. However, scalability remains a significant challenge when addressing large-scale, real-world freight networks with thousands of nodes, multiple transportation modes, and complex operational constraints. While small instances can be solved to optimality using exact methods like mixed-integer programming in Gurobi, larger problem instances may require computationally intensive approaches that exceed reasonable time limits for real-time decision-making. For such scenarios, the system must resort to efficient heuristic methods, metaheuristics, or approximation algorithms, which may compromise solution quality in favor of computational tractability. This trade-off between solution optimality and computational efficiency represents a fundamental limitation in scaling the framework to metropolitan or national-level freight networks.

\textbf{Natural Language Understanding and Domain Contextualization:} Although the chatbot interface improves accessibility for non-technical users, natural language understanding and scenario interpretation remain constrained by the language model's coverage of domain-specific context. Misinterpretation of ambiguous queries, incomplete specification of constraints, or over-reliance on pre-trained embeddings may result in suboptimal tool selection or workflow planning. The current system's ability to understand nuanced logistics terminology and translate user intentions into precise mathematical formulations requires further refinement.

Future research directions should address these limitations through several key initiatives: developing standardized ontologies and data exchange protocols for freight transportation tools; creating automated data preprocessing pipelines that can handle diverse data formats; implementing adaptive optimization strategies that dynamically select between exact and heuristic methods based on problem complexity; establishing real-time data integration capabilities through IoT networks and live traffic feeds; and expanding domain coverage to support broader urban logistics applications beyond freight transportation.

\section{{CONCLUSION}}
This paper presents a novel framework that integrates generative AI, multi-agent systems, and the MCP to enable autonomous, simulation-informed decision-making in urban freight logistics. By combining large language models with domain-specific optimization and simulation tools such as Gurobi and AnyLogic, the proposed system represents a fundamental paradigm shift—transforming digital twins from static, visual representations into dynamic, intelligent agents capable of autonomous reasoning and decision-making.

The framework successfully demonstrates the integration of retrieval-augmented generation, structured knowledge graphs, and microservice-based orchestration to empower the system to interpret user-defined objectives, generate actionable workflows, and deliver optimized solutions through natural language interaction. Through comprehensive use cases, we have shown how the framework enables users—regardless of technical background—to engage with complex logistics problems, receive optimized intermodal routing recommendations, and visualize operational strategies in real time. The system's ability to autonomously manage tool chains and interpret scientific outputs in human-readable form significantly reduces barriers to accessing advanced decision support capabilities in practice.

It is noteworthy that when this research was initiated, the MCP—now recognized as a critical enabler for AI agent orchestration—had not yet gained widespread attention in the research community. The rapid evolution of generative AI technologies and multi-agent systems has transformed the landscape of intelligent software development, making previously theoretical concepts practically implementable. Our work represents an early exploration into this emerging paradigm, demonstrating the feasibility of autonomous scientific workflow orchestration through AI agents. While the field continues to evolve rapidly, with new standards, protocols, and capabilities emerging frequently, we have established a foundational framework that can adapt to and incorporate future technological advances.

The proposed framework works effectively within its current scope, successfully addressing the core challenges of knowledge integration, tool orchestration, and user accessibility in urban freight optimization. However, significant opportunities remain for enhancement and expansion. Future developments will focus on addressing scalability challenges, improving data integration capabilities, standardizing tool interoperability protocols, and expanding domain applicability beyond freight transportation to encompass broader urban systems management.

Overall, this work advances the frontier of digital twin applications in urban operations research by offering a flexible, scalable, and user-friendly platform for supporting data-driven, low-carbon freight transportation planning. As generative AI technologies continue to mature and standardization efforts like MCP gain broader adoption, we anticipate that intelligent, autonomous digital twins will become essential tools for addressing complex urban challenges. Our contribution represents a significant first step toward realizing this vision, establishing both the technical foundation and practical methodology for developing truly intelligent urban management systems.

\section*{Acknowledgments}
This work was supported in part by the U.S. Department of Energy (DOE), Advanced Research Projects Agency–Energy (ARPA-E), under project DE-AR0001780. We extend our gratitude to our collaborators from the University of Tennessee, Knoxville, the Oak Ridge National Laboratory, and West Virginia University. Additionally, several icons from www.flaticon.com were utilized in creating figures for this research.

\section*{Generative Artificial Intelligence (AI) Statement}
As part of this research, we used Generative Artificial Intelligence tools to support content development, including grammar correction and LaTeX formatting. Specifically, we employed ChatGPT-4o, Ollama 3.1 (8B), and Claude Sonnet 4 to assist in editing and document preparation.

\section*{Disclaimer}
This manuscript has been authored by UT-Battelle, LLC, under contract DE-AC05-00OR22725 with the US Department of Energy (DOE). The US government retains and the publisher, by accepting the article for publication, acknowledges that the US government retains a nonexclusive, paid-up, irrevocable, worldwide license to publish or reproduce the published form of this manuscript, or allow others to do so, for US government purposes. DOE will provide public access to these results of federally sponsored research in accordance with the DOE Public Access Plan (\url{http://energy.gov/downloads/doe-public-access-plan}).

\bibliographystyle{apalike}
\bibliography{refs}

\end{document}